\documentclass[prl,footinbib,showpacs,twocolumn,amsmath,amssymb]{revtex4}
\usepackage{graphicx}%
\usepackage{xcolor}
\usepackage{bm}
\usepackage{amsmath, amsthm, amssymb}
\usepackage{braket}
\renewcommand\Im{\operatorname{Im}}

\usepackage[normalem]{ulem}
\usepackage{upgreek}

\makeatletter
\newcommand\footnoteref[1]{\protected@xdef\@thefnmark{\ref{#1}}\@footnotemark}
\makeatother

\definecolor{orange}{rgb}{1,0.5,0}
\usepackage{scrextend}

\begin{document}

\title{Tuning paramagnetic spin-excitations of single adatoms}

\author{Julen Iba\~{n}ez-Azpiroz, Manuel dos Santos Dias, Benedikt Schweflinghaus, 
Stefan Bl\"ugel, Samir Lounis}
\address{Peter Gr\"unberg Institute and Institute for Advanced Simulation, Forschungszentrum
J\"ulich \& JARA, D-52425 J\"ulich, Germany}
\date{\today}


\begin{abstract}

We predict the existence of paramagnetic spin-excitations (PSE) in non-magnetic single adatoms.  
Our calculations demonstrate that PSE
develop a well-defined structure in the meV region when the adatom's
Stoner criterion for magnetism is close to the critical point. 
We further reveal 
a subtle tunability and enhancement of PSE by external magnetic fields.  
Finally, we show how PSE can be 
detected as moving steps in the $\mathrm{d}I/\mathrm{d}V$ signal of   
inelastic scanning tunneling spectroscopy,
opening a potential route for experimentally accessing electronic 
properties of non-magnetic adatoms, such as the Stoner parameter.

\end{abstract}
\maketitle

Single adatoms deposited on surfaces have 
become a prominent  playground where 
theory and experiment can explore hand by hand 
a large variety of physical phenomena
ranging from spin-excitations
~\cite{hirjibehedin_large_2007,PhysRevLett.102.256802,PhysRevLett.103.176601,lounis_dynamical_2010,zhou_strength_2010,PhysRevLett.106.037205,PhysRevLett.111.157204,ternes_spin_2015,khajetoorians_tailoring_2016,ibanez-azpiroz_zero-point_2016} to 
magnetic exchange interactions~\cite{oberg_control_2014,yan_control_2015,PhysRevLett.107.186805},
quantum spin decoherence~\cite{bryant_controlled_2015,baumann_electron_2015,delgado_spin_2017},
topological superconductivity~\cite{nadj-perge_observation_2014,PhysRevLett.111.147202,PhysRevLett.111.206802}
or the Kondo effect~\cite{ternes_spectroscopic_2009,PhysRevLett.114.076601},
among many others. 
Virtually all these effects arise from the 
intricate interplay between
the degrees of freedom of the adatom - charge, spin or orbital momentum -
and 
the electron and phonon bath of the substrate, 
a subject of heavy and ongoing investigation. 

Noteworthily, magnetism plays a central role in fueling the interest for single adatoms,
given that they represent the ultimate limit
in the context of bit miniaturization in data storage devices. 
As a consequence, great efforts are being devoted to the search and characterization
of elements that become magnetic when deposited on a substrate.
Successful examples include, \textit{e.g.},
Fe and Co on Pt(111)~\cite{gambardella_giant_2003,PhysRevLett.111.157204},
Fe on Cu(111)~\cite{PhysRevLett.106.037205} as well as on Cu$_{2}$Ni/Cu(111)~\cite{PhysRevLett.111.127203}
and CuNi~\cite{hirjibehedin_large_2007},
Co on MgO(100)~\cite{rau_reaching_2014},
and more recently
Ho on MgO/Ag(100)~\cite{donati_magnetic_2016}, 
which all  
exhibit local magnetic moments greater than $2$ $\mu_{B}$ and reveal 
clear signatures of magnetism that manifest either in 
a large magnetic anisotropy energy, steps in the $\mathrm{d}I/\mathrm{d}V$ signal 
related to spin-excitations or even remanence of the magnetic signal.

In this Letter, we propose and argue that 
even nominally non-magnetic single adatoms can 
exhibit clear fingerprints of magnetism in the form of well-defined 
features in the spin-excitation spectrum, 
\textit{i.e.}, paramagnetic spin-excitations (PSE). 
Interestingly, these are the analogous of so-called paramagnons 
first proposed by Doniach in 1967~\cite{doniach_theory_1967}
and first measured in bulk Pd 
nearly 50 years later by Doubble \textit{et. al.}~\cite{PhysRevLett.105.027207}
(see also Ref. \onlinecite{staunton_spin_2000} for recent calculations). 
In the context of Fermi liquid theory, these excitations can be viewed as
persistent spin-fluctuation modes that can be activated by temperature and 
thus produce a measurable impact on 
properties such as specific heat or electron 
effective-mass enhancement~\cite{doniach_theory_1967,lonzarich_effect_1985}.
Upon reducing the dimensionality of the system, here we show that PSE can be 
strongly enhanced due to the modified interplay between the two 
fundamental electronic properties involved, namely the 
Stoner exchange interaction and the adatom's density of states (DOS)
at the Fermi level.
Importantly, this opens up unforeseen potential applications  
of non-magnetic adatoms in nanotechnology, 
which encodes and manipulates information  
into excitation modes like PSE.
In addition, our \textit{ab-initio} analysis based on time-dependent density functional theory
(TDDFT) reveals that PSE are highly 
sensitive to externally applied magnetic fields 
and, furthermore, can exhibit a singular enhancement 
when the field approaches a critical regime.
Motivated by these findings, we assess the impact of PSE on the 
$\mathrm{d}I/\mathrm{d}V$ signal as measured in
inelastic scanning tunneling spectroscopy (ISTS) experiments,
identifying clear signatures of magnetic response 
that allow to distinguish these type of excitations from, 
\textit{e.g.}, phonons.

A central property for our discussion is the spin-excitation spectrum of non-magnetic adatoms.
Within the TDDFT formalism, this information is encoded into the 
longitudinal component of the enhanced spin-susceptibility, $\chi(\omega)$, which is related 
to the response of the non-interacting Kohn-Sham (KS) system, $\chi^{KS}(\omega)$:~\cite{aguayo_why_2004}
\begin{equation}\label{eq:chi-general}
\chi(\omega)=\dfrac{\chi^{KS}(\omega)}{1-I_{s}\chi^{KS}(\omega)}.
\end{equation}
Above, $I_{s}$ denotes the so-called Stoner parameter,
which plays the role of the exchange-correlation kernel in the 
adiabatic local spin-density approximation~\cite{vosko_accurate_1980}.
Noteworthily, the static limit of Eq. \ref{eq:chi-general} 
recovers the standard Stoner theory that provides the well-known criterion
for magnetism, \textit{i.e.} $\chi(0)<0\Rightarrow I_{s}\rho_{F}>1$, with 
$\rho_{F}$ the adatom's DOS at the Fermi level  and we used $\chi^{KS}(0)=\rho_{F}>0$~\cite{aguayo_why_2004}.
In essence, the product $I_{s}\rho_{F}$ quantifies the competition
between the exchange interaction, which  
enhances the tendency towards magnetism of electrons in localized orbitals, 
and substrate hybridization, which induces delocalization of the adatom's electrons and 
therefore acts against magnetism, thus playing the role of 
the kinetic energy in the standard Stoner theory.
It is interesting to note that 
even if an adatom does not fulfill the Stoner criterion,
it can still develop dynamical PSE provided the details of the electronic structure make the denominator
of Eq. \ref{eq:chi-general} vanishingly small at a finite frequency.

\begin{figure}[t]
\includegraphics[width=0.5\textwidth]{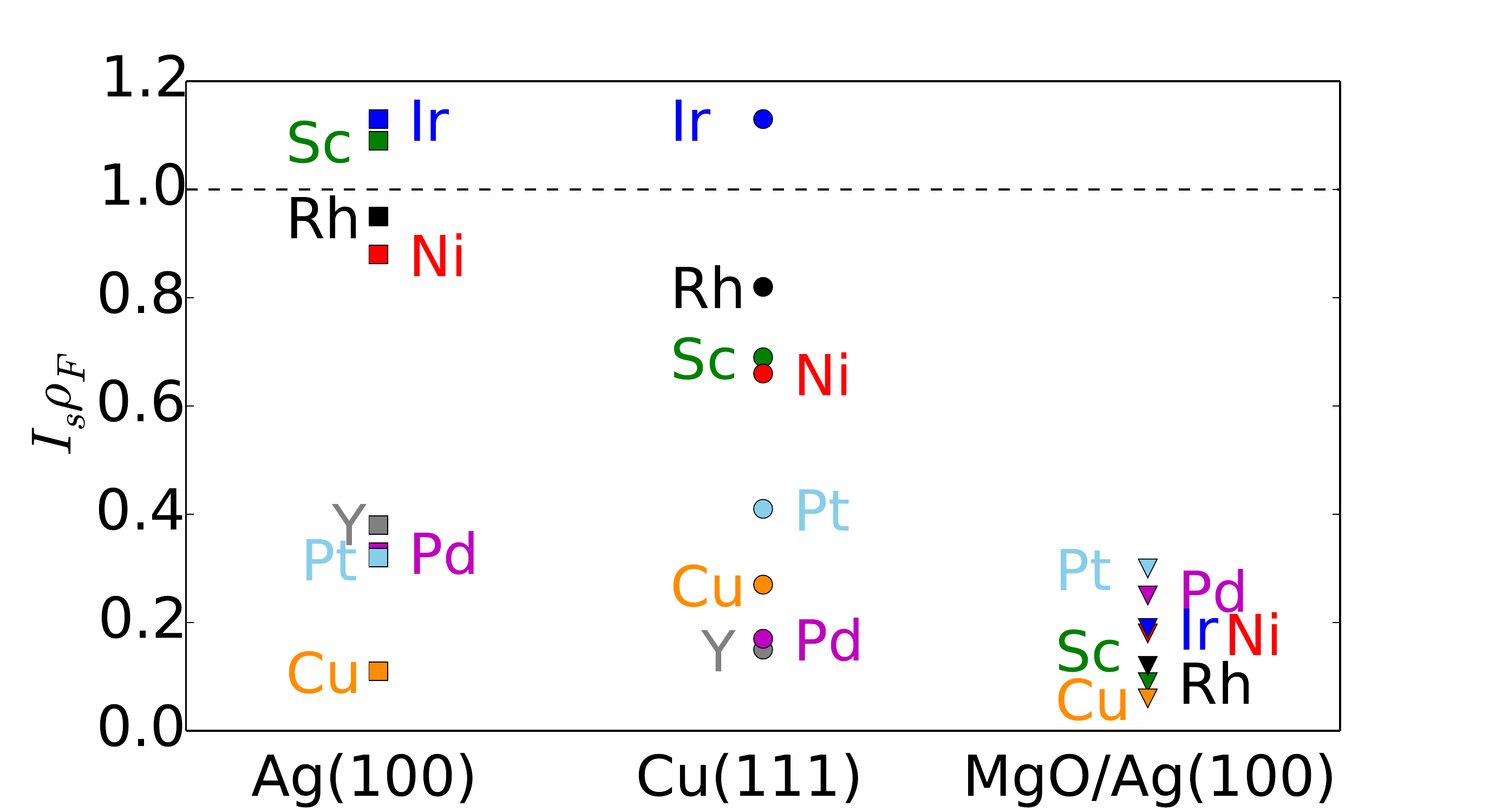}
\caption{(color online) Calculated Stoner product for various \textit{3d, 4d} and \textit{5d}
transition metal adatoms deposited on Ag(100) (squares), Cu(111) (circles) and MgO/Ag(100) (triangles).
}
\label{fig:IS-table}
\end{figure}

Let us begin our analysis by characterizing the set of \textit{3d, 4d} and \textit{5d} 
transition metal adatoms
that could potentially exhibit PSE. 
For this purpose, in Fig. \ref{fig:IS-table}
we list several adatoms whose calculated Stoner products 
are below or slightly above 1; the calculations have been
performed following the Korringa-Kohn-Rostoker Green function  
formalism~\cite{papanikolaou_conceptual_2002,lounis_dynamical_2010,lounis_theory_2011}
(see Supplemental Material for technical details, which includes 
Refs. 
\onlinecite{espresso,janak_uniform_1977,PhysRevB.65.104441,beckmann_magnetism_1997,PhysRevB.52.11502}) and 
considering three different substrates, namely  Ag(100), Cu(111)
and MgO/Ag(100).  
As a general trend, our calculations show that the metallic substrates Ag(100) and Cu(111) host adatoms
whose Stoner product is closer to the critical value 1 as compared to insulating MgO/Ag(100).
This is mainly due to the small $\rho_{F}$ in the later,
as tabulated in the Supplemental Material.
Among the two metallic substrates, Ag(100)  
hosts adatoms whose Stoner product are closest to 1, 
with $I_{s}\rho_{F}$ ranging between $\sim[1-0.1,1+0.1]$ for Sc, Ir, Rh and Ni adatoms.
Therefore, throughout the work we will focus on discussing the Ag(100) substrate in detail,
as it illustrates best our findings.

\begin{figure}[b]
\includegraphics[width=0.5\textwidth]{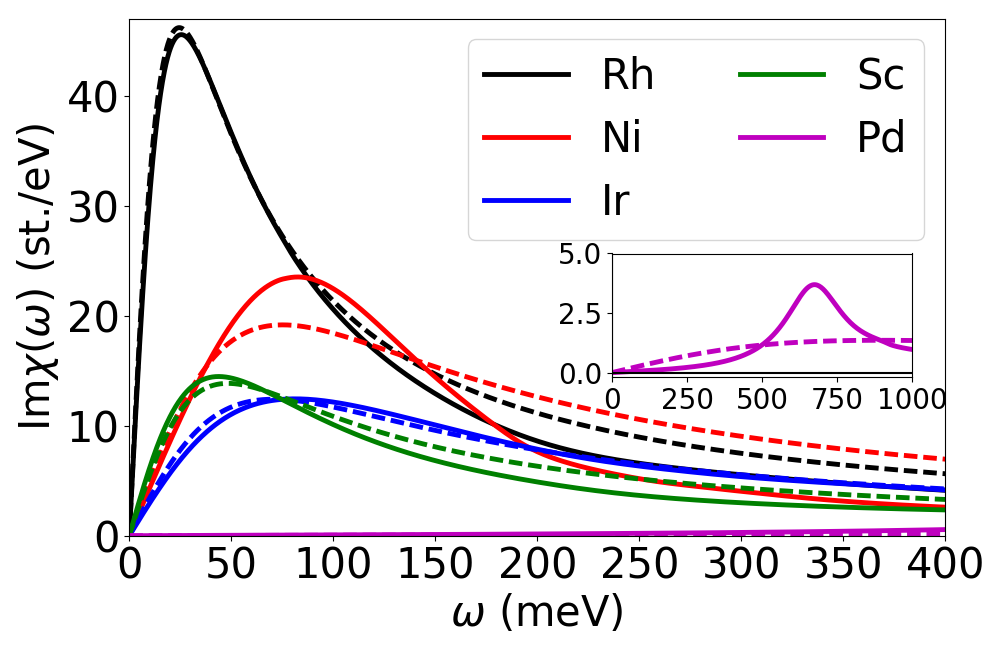}
\caption{(color online) Solid lines illustrate the calculated 
density of PSE as given by 
$\Im\chi(\omega)$  (Eq. (\ref{eq:chi-general})) 
of selected \textit{3d, 4d} and \textit{5d} transition metal 
adatoms deposited on the metallic Ag(100) substrate. 
Dashed lines denote the approximation of Eq. (\ref{eq:imchi-linear-w}).
Note that both Eqs. (\ref{eq:chi-general}) and (\ref{eq:imchi-linear-w})
give rise to PSE.
}
\label{fig:PSE}
\end{figure}

In Fig. \ref{fig:PSE} we illustrate the  calculated spin-excitation spectra as given by 
$\Im\chi(\omega)$ from Eq. (\ref{eq:chi-general}),
where all calculations were done considering the non-magnetic ground state
(see Supplemental Material for technical details).  
Interestingly, Fig. \ref{fig:PSE} reveals peak-like structures resonating at frequencies below 100 meV
for Rh, Ni, Ir and Sc adatoms.
This is exceptional, as most non-magnetic elements 
exhibit a featureless spectrum owing to a complete overdamping of the excitations.
Rh represents the most favorable case, displaying
a well-defined peak at $\omega_{\text{res}}\sim 20$ meV 
and a width of 
$\Delta\sim50$ meV,
the associated lifetime being $\tau=\omega_{\text{res}}^{-1}\sim30$ fs.
It is noteworthy that both the lifetime and the height of the peak, 
the later being related to the intensity of the excitation, are only one order
of magnitude smaller than those of usual transverse spin-excitations
measured by ISTS in magnetic adatoms, such as Fe on Cu(111)
(see, \textit{e.g.}, Refs. \onlinecite{PhysRevLett.106.037205,dias_relativistic_2015}).
On the other extreme, Pd in Fig.  \ref{fig:PSE} shows a highly overdamped resonance at around
600 meV (see figure inset) whose intensity is an order of magnitude smaller than that of Rh.
Therefore, our \textit{ab-initio} calculations reveal the existence of PSE 
whose resonance frequency and width vary strongly depending on the adatom.

Next, we focus on characterizing the physical mechanism behind PSE
that allows an interpretation of the \textit{ab-initio} results displayed in Fig. \ref{fig:PSE}.
For this purpose, let us consider the frequency expansion of the paramagnetic KS spin response function
up to linear order, \textit{i.e.}, 
$\chi^{KS}(\omega)=\rho_{F}+i\alpha\omega+ \mathcal{O}(\omega^{2})$.
One can show (see Supplemental Material) that the linear expansion
coefficient is well approximated by $\alpha\sim -\pi \rho^{2}_{F}/4$.
Therefore, the spin-excitation spectrum within this approximation 
is given by a simple expression involving only the DOS at $E_{F}$ and the Stoner parameter:
\begin{equation}
\label{eq:imchi-linear-w}
\Im\chi(\omega)= \dfrac{\pi}{4}\dfrac{\rho^{2}_{F}\omega}{\big(1-I_{s}\rho_{F} \big)^{2}
+(\frac{\pi}{4}I_{s}\rho^{2}_{F} \omega)^{2}}.
\end{equation}
By extracting $\rho_{F}$ and $I_{s}$ from our \textit{ab-initio} calculations, we have 
computed and displayed the expression predicted by Eq.~(\ref{eq:imchi-linear-w}) 
for each of the adatoms
considered in Fig. \ref{fig:PSE} (see dashed lines). 
A comparison to the full \textit{ab-initio} calculations (solid lines)
reveals a very good agreement for frequencies below 100 meV 
in the case of Rh, Ir and Sc, where both the peak and width are 
properly described within $\leqslant 10\%$ relative error.
This error is considerably larger in the case of Ni, indicating the importance
of higher order expansion terms in $\omega$ for this case. 
Finally, the peak for Pd is far beyond the limit of small frequencies and therefore
the approximation of Eq.~(\ref{eq:imchi-linear-w}) breaks down.

\begin{figure*}[t]
\includegraphics[width=1.0\linewidth]{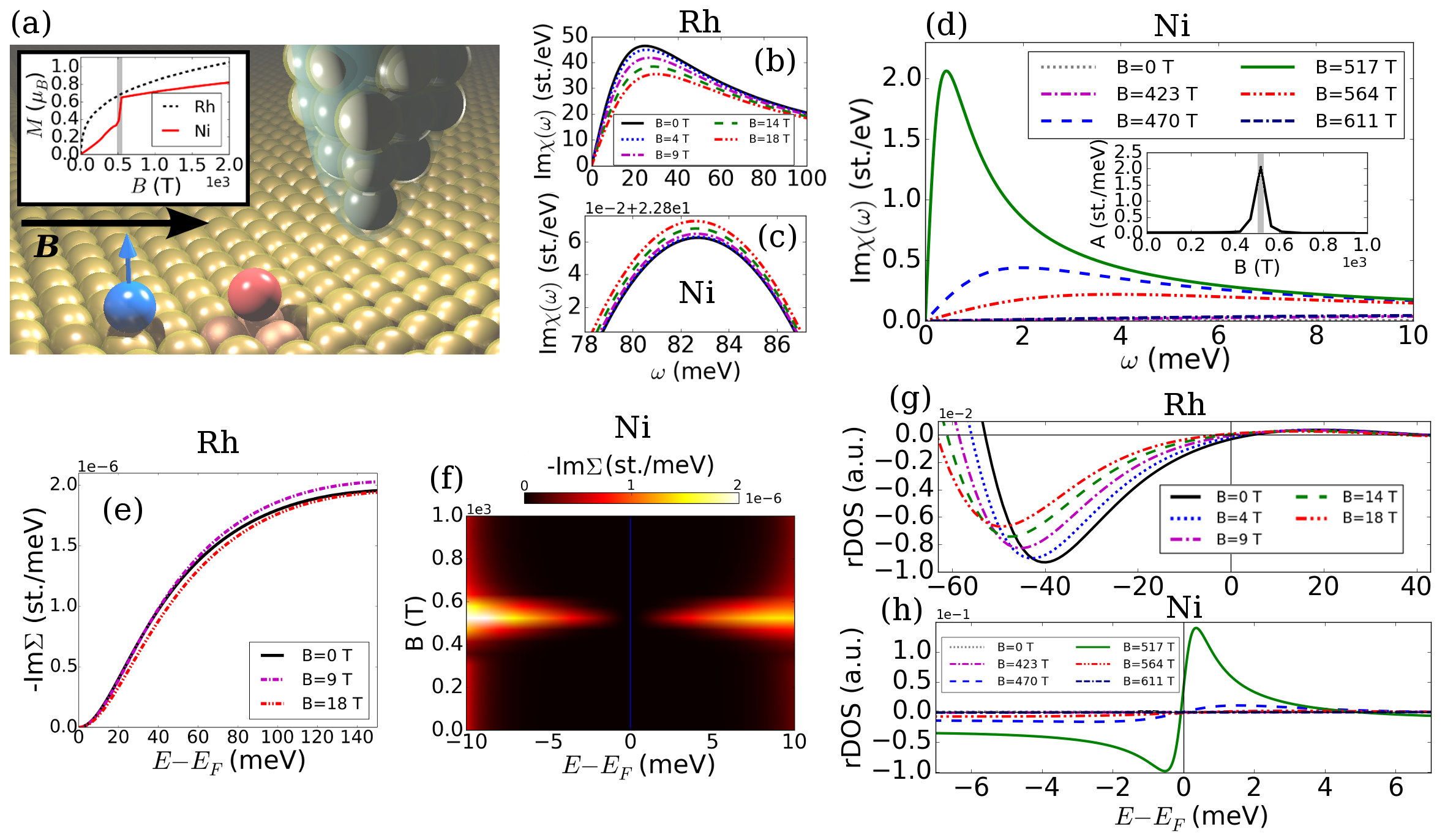}
\caption{(color online) (a) Minimal setup illustrating the proposed ISTS measurement. 
Substrate, non-magnetic adatom and tip atoms are displayed as 
gold, red and grey balls, respectively, while the black arrow
depicts an external magnetic field $B$.
The graph in the inset illustrates the calculated magnetic moment $M$ as a function
of the external field for Rh and Ni adatoms, with the grey area indicating the
critical regime of Ni where $M$ shows a discontinuity. 
A blue ball with an arrow has been added in the main figure to illustrate the possibility of 
coupling a magnetic adatom to the non-magnetic one, inducing on the later a
magnetic moment of the order of the values shown in the 
inset, thus mimicking the effect of large magnetic fields~\cite{footnote}.
The rest of subfigures show the calculated magnetic field dependence of various properties.
(b) and (c) Density of PSE as given by 
$\Im\chi(\omega)$ for Rh and Ni adatoms, respectively, for magnetic fields of up to 18 T
(both figures share the same legend).
(d) Same as in (c) but for larger magnetic fields of up to 10$^{3}$ T. The inset 
depicts the evolution of the PSE's amplitude (see Eq.~(\ref{eq:resonance-freq})) 
as a function of the magnetic field.
(e) and (f) Imaginary part of the 
self-energy, $\Im\Sigma(V_{F})$, for
Rh and Ni adatoms, respectively. Note the different 
scale of the magnetic field and the energy window in the two cases.
Vertical (blue) line in (f) separates negative and positive energies.
(g) and (h) Energy derivative of the renormalized DOS (s orbital)
for Rh and Ni, respectively.
Note the difference in magnitude on the applied magnetic fields in both cases.
}
\label{fig:STM}
\end{figure*} 

Proving Eq.~(\ref{eq:imchi-linear-w}) to be an accurate approximation 
of the full spin-excitation density given by Eq. (\ref{eq:chi-general}) is extremely convenient, 
as the former provides an analytical interpretation for the origin of PSE 
in terms of just $\rho_{F}$ and $I_{s}$, two basic electronic properties of adatoms.
Indeed, the resonance frequency, linewidth and 
amplitude of PSE 
predicted by Eq.~(\ref{eq:imchi-linear-w}) can be cast into simple expressions:
\begin{equation}
\label{eq:resonance-freq}
\begin{aligned}
&\omega_{\text{res}}=\dfrac{4}{\pi}\frac{|1-I_{s}\rho_{F}|}{I_{s}\rho^{2}_{F}},\;\; 
\Delta = 2\sqrt{3}\omega_{\text{res}},\\
& \;\; A\equiv \Im\chi(\omega_{\text{res}})=\dfrac{1}{2I_{s}|1-I_{s}\rho_{F}|} .
\end{aligned}
\end{equation}
Interestingly, a potential  measurement  of the above quantities
would directly yield experimental estimates for $\rho_{F}$ and $I_{s}$.
In closer inspection, one recognizes the Stoner product $I_{s}\rho_{F}$ 
as the key quantity in Eq.~(\ref{eq:resonance-freq}); 
as $I_{s}\rho_{F}\rightarrow 1$ (\textit{i.e.}, ferromagnetic instability),
the resonance frequency as well as the linewidth tend 
to zero while the intensity of PSE
shows a singularity. 
This analysis offers therefore the interpretation we seeked for, namely 
that elements closer to the ferromagnetic instability 
show enhanced PSE, as it can be clearly checked from the comparison of
Figs. \ref{fig:IS-table} and \ref{fig:PSE}. 
We emphasize that the mechanism just described is 
fundamentally different from the one taking place in magnetic adatoms, 
where the resonance frequency of transverse spin-excitations 
is settled by the spin-orbit 
interaction via the magnetic anisotropy energy~\cite{dias_relativistic_2015}.

Having exposed the origin of PSE in single-adatoms, 
we focus next on assessing their potential impact on the
$\mathrm{d}I/\mathrm{d}V$ signal as measured in ISTS experiments,
the technique of choice for measuring magnetic excitations 
(see, \textit{e.g.}, 
Refs.~\onlinecite{PhysRevLett.106.037205,PhysRevLett.111.157204,steinbrecher_absence_2016}).
The corresponding minimal setup is illustrated in Fig. \ref{fig:STM}(a),
which displays a scanning tunneling microscope (STM) tip 
measuring the adatom's excitations under an applied external magnetic field, denoted
as $B$. 
We first notice that
PSE respond to magnetic fields by shifting their resonance frequency.
This is quantitavely demonstrated in Figs. \ref{fig:STM}(b) and \ref{fig:STM}(c), 
where the calculated spin-excitation spectra 
are shown for Rh and Ni adatoms, respectively, 
under $B$ fields of $\sim$10 T that are achieveable in state-of-the-art laboratories
(see, \textit{e.g.}, Refs. \onlinecite{PhysRevLett.111.157204,rau_reaching_2014,donati_magnetic_2016}). 
Noteworthily, while the PSE of Rh 
shifts towards larger frequencies as $B$ is increased (see Fig. \ref{fig:STM}(b)), 
the PSE of Ni exhibits the opposite behavior (see Fig. \ref{fig:STM}(c)).
This difference arises from the fact that magnetic fields 
induce an effective modification of Stoner product, \textit{i.e.}, 
$I_{s} \rho_{F}\rightarrow \xi(B)I_{s} \rho_{F}$,
where $\xi(B)$ is a term that depends both on the magnetic field 
as well as on the adatom's electronic structure (see Supplemental Material).
In particular, the details of the later make $\xi(B)>1$ for Ni while 
$\xi(B)<1$ for Rh, leading to the aforementioned divergent responses
in accordance with Eq.~(\ref{eq:resonance-freq}).

Remarkably, when strong enough magnetic fields are applied to Ni, 
the modified Stoner criterion
can be tuned towards the critical point, 
as shown in Fig. \ref{fig:STM}(d).
As a consequence, the PSE's resonance frequency approaches the origin in a singular way
while the amplitude of the excitation is enhanced by as much as two orders of magnitude for 
$B\sim 500$ T.  
It is interesting to note that this critical behavior 
is also present on the $B$-field dependence of the induced magnetic moment $M$, 
as shown in the inset of Fig. \ref{fig:STM}(a).
While Rh shows a continuous dependence, Ni reveals a discontinous transition 
at approximately the critical field value $B\sim 500$ T, above which
the system enters a magnetic regime where 
the internal exchange field effectively contributes to  $M$ on top of the
external Zeeman field, featuring the atomic version of
a quantum phase transition.
We note that, although such large $B$ fields 
are clearly out of reach for current experiments,
this feature could be potentially observed,  
\textit{e.g.}, 
via the proximity effect, by placing a 
magnetic adatom in the neighborhood of the non-magnetic one
(see Fig. \ref{fig:STM}(a)).
Our calculations verify that the former can induce on the later a magnetic moment of the
same order of magnitude as the one induced by the fields 
of Fig. \ref{fig:STM}(d)~\cite{footnote},
thus mimicking the action of large magnetic fields.

Next we evaluate the impact of PSE on the $\mathrm{d}I/\mathrm{d}V$ signal
of an ISTS measurement. For such purpose  we consider the so-called Tersoff-Hamann
approximation~\cite{PhysRevLett.50.1998,PhysRevLett.86.4132}, which relates the ISTS spectrum to 
the electronic DOS at the tip position renormalized by the adatom's excitations. 
We access the latter quantity by means of a recently developed technique
that combines many body perturbation theory with our TDDFT scheme; details
can be found in Ref. \onlinecite{PhysRevB.89.235439}.
The central object within this formalism is the electron self-energy, 
$\Sigma$, which contains 
the interactions between the tunneling electrons from the tip at bias voltage $V$
and the adatom's PSE. It is particularly revealing
to inspect its imaginary part~\cite{PhysRevB.89.235439},  
\begin{equation}
\label{eq:selfe}
\begin{aligned}
&\Im\Sigma(V_{F}) = - I_{s}^{2}\int_{0}^{-V} d\omega \; \rho(V_{F}+\omega) \Im\chi(\omega),
\end{aligned}
\end{equation}
with $\rho(E)$ the energy-dependent DOS, $V_{F}=E_{F}+V$ and $E_{F}$
the Fermi energy.  The calculated $\Im\Sigma(V_{F})$  is shown in Fig.  \ref{fig:STM}(e) 
for Rh under various magnetic fields of up to 18 T.  
Our results reveal a clear step for positive bias voltage 
that saturates at $\sim 100$ meV, \textit{i.e.}, after the PSE peak has been integrated 
(see Eq. (\ref{eq:selfe})). 
Note also that the calculated self-energy
slightly varies as a function of the magnetic field.
When larger magnetic fields are applied, as illustrated in Fig. \ref{fig:STM}(f) for the case of Ni,
the critical behavior of the PSE (see Fig. \ref{fig:STM}(d)) translates into  a
clear maximum at the value of the critical field,
where $\Im\Sigma(V_{F})$ increases by an order of magnitude.

The presence of PSE has a broad effect on the 
renormalization of the DOS at the vacuum, where
ISTS tips measure the signal. In particular, the energy derivative of the 
renormalized DOS (rDOS) is a quantity that is linked to the $\mathrm{d}^{2}I/\mathrm{d}V^{2}$ curve
measured by ISTS~\cite{PhysRevB.89.235439}.  
The former quantity is displayed in Fig. \ref{fig:STM}(g) for Rh, 
where the magnetic field dependence is clearly visible. 
Noteworthily, our calculations demonstrate that 
the tunneling electrons from the tip are able to trigger the PSE,
leading to a peak in the meV region that, furthermore, 
reacts to external magnetic fields
by shifting its resonance frequency as well as 
substantially modifying its intensity. 
We also note the strong asymmetric distribution 
between positive and negative frequencies, a feature that  emerges from the
background electronic structure~\cite{PhysRevB.89.235439} and is 
commonly present in  $\mathrm{d}^{2}I/\mathrm{d}V^{2}$ curves measured on magnetic adatoms  
(see, \textit{e.g.}, 
Refs.~\cite{heinrich_single-atom_2004,PhysRevLett.106.037205,PhysRevLett.110.157206,PhysRevLett.111.157204,PhysRevLett.115.237202}).
On the other hand, when Ni is driven into the critical regime as in Fig. \ref{fig:STM}(h),
our calculations reveal a huge change of the signal's intensity 
as the PSE approaches the critical point. 
Our analysis therefore shows that magnetism offers 
a prime way of manipulating PSE, enabling to discern them from other
excitations of similar energy but non-magnetic origin, such as phonons.

In conclusion, we have proposed and argued a means of detecting  
spin-excitations in non-magnetic single adatoms.
We have shown that such excitations can develop well defined peaks in the meV
region, their main characteristics
being determined by two fundamental electronic properties, namely the 
Stoner parameter and the DOS at the Fermi level.
Our analysis based on TDDFT has further revealed a pronounced dependence 
of PSE on externally applied magnetic fields, 
exhibiting 
the atomic analogue of a quantum phase transition
as the field approaches the critical value. 
This remarkable feature is likely to have strong effects
in processes where a substantial magnetic moment is induced in
non-magnetic adatoms, \textit{e.g.}, when magnetic atoms are 
coupled to them via the proximity effect. 
Finally, we have simulated \textit{ab initio} the impact 
of PSE on the $\mathrm{d}^{2}I/\mathrm{d}V^{2}$ curve measured in state-of-the-art ISTS experiments,
revealing that PSE can be triggered by tunneling electrons 
and, furthermore, exhibit a clear response to magnetic fields.
Thus, besides opening up potential applications for non-magnetic adatoms, 
our analysis offers a
route for experimentally accessing their fundamental electronic 
properties, 
such as the Stoner parameter.

This work has been supported by the
Helmholtz Gemeinschaft Deutscher-Young Investigators
Group Program No. VH-NG-717 (Functional Nanoscale
Structure and Probe Simulation Laboratory), the Impuls und
Vernetzungsfonds der Helmholtz-Gemeinschaft Postdoc Programme, 
and funding from the European Research Council (ERC) under 
the European Union's Horizon 2020 research and innovation 
programme (ERC-consolidator grant 681405 — DYNASORE)

\textit{Note added in Proof} ---
In the recent work of Ref.~\onlinecite{schendel_strong_2017},
the conductance associated to a single Pd adatom deposited
on Pd(111) has been experimentally measured and interpreted 
as being strongly affected by paramagnon scattering.

\section*{References}
\bibliography{biblio}

\begin{thebibliography}{48}
\bibitem[{foo()}]{footnote}
\bibinfo{note}{For externally applied $B= 517$ T the magnetic moment of Ni is
  $m=0.29$ $\mu_{B}$, while a Fe adatom placed as a second nearest neighbor
  induces a magnetic moment on Ni of $m=0.51$ $\mu_{B}$.}
\end{thebibliography}

\end{document}